# HOW CONTENT VOLUME ON LANDING PAGES INFLUENCES CONSUMER BEHAVIOR: EMPIRICAL EVIDENCE

| | | |
|---|---|---|
| Ruti Gafni* | The Academic College of Tel Aviv Yaffo, Tel Aviv, Israel | rutigafn@mta.ac.il |
| Nim Dvir | University at Albany, State University of New York, Albany, NY, USA | Ndvir@albany.edu |

* Corresponding author

## ABSTRACT

| | |
|---|---|
| Aim/Purpose | This paper describes an empirical investigation on how consumer behavior is influenced by the volume of content on a commercial landing page -- a stand-alone web page designed to collect user data (in this case the user's e-mail address), a behavior called "conversion." |
| Background | Content is a term commonly used to describe the information made available by a website or other electronic medium. A pertinent debate among scholars and practitioners relate to information volume and consumer behavior: do more details elicit engagement and compliance, operationalized through conversions, or the other way around? |
| Methodology | A pilot study (n= 535) was conducted in real-world commercial setting, followed by a series of large-scale online experiments (n= 27,083). Both studies employed a between-group design: Two variations of landing pages, long and short, were created based on various behavioral theories. User traffic to the pages was generated using online advertising and randomized between the pages (A/B testing). |
| Contribution | This research contributes to the body of knowledge on the antecedents and outcomes of online commercial interaction, focusing on content as a determinant of consumer decision-making and behavior. |
| Findings | The observed results indicate a negative correlation between content volume and users' conversions. The shorter pages had significantly higher conversion rates, across locations and time. Findings suggest that content play a significant role in online decision making. They also contradict prior research on trust, persuasion, and security. |
| Recommendations for Practitioners | At a practical level, results can inform practitioners on the importance of content in online commerce. They provide an empirical support to design and content strategy considerations, specifically the use of elaboration in commercial web pages. |
| Recommendation for Researchers | At the theoretical level, this research advances the body of knowledge on the paradoxical relationship between the increased level of information and online decision-making and indicates that contrary to earlier work, not all persuasion theories are effective online. |
| Impact on Society | Understanding how information drive behavior has implications in many domains (civic engagement, health, education and more). This has relevance to system design and public communication in both online and offline contexts, suggesting social value. |
| Future Research | Using this research as a starting point, future research can examine the impact of content in other contexts, as well as other behavioral drivers (such as demographic data). This can lead to theoretical, methodological and practical recommendations. |
| Keywords | content, landing pages, content strategy, impression-management, decision-making, human-computer interaction, engagement, a/b testing, e-commerce, marketing |

## INTRODUCTION

This paper explores the impact of digital content on consumer behavior in online commercial settings. Digital content is defined as the textual or visual information made available by a website or other electronic medium (Gates, 1996; Huizingh, 2000; Rowley, 2008). In marketing and online commerce, digital content and digital information are often synonymous terms (Rowley, 2008). Digital content has been proven to play a significant role in influence consumers' behavior online (Jarvenpaa, Tractinsky, & Saarinen, 2006; Jennings, 2000; Lee & Turban, 2001; Wells, Valacich, & Hess, 2011). This is particularly true regarding landing pages -- a single web page that appears in response to clicking on an online advertisement and aims to persuade a visitor to take action by completing a transaction, usually by providing a form that needs to be filled out (Becker, Broder, Gabrilovich, Josifovski, & Pang, 2009; Unbounce, 2016). This action performed by the user is called "conversion" or "compliance".

As landing pages are often users' first impression of a web page, decisions are greatly based on the content being presented (Ash, Page, & Ginty, 2012; Becker et al., 2009; Lindgaard, Fernandes, Dudek, & Brown, 2006; Reinecke et al., 2013). This study empirically addresses a long-standing question: Does more content (information volume) elicit conversion or the other way around? Despite decades of research, there is still a conceptual confusion on the topic. A pertinent debate among scholars and practitioners relates to the question of content volume (sometimes called "amount") versus user behavior. While there seems to be a consensus that content impact behavior, there is no agreement on volume of content: On the one hand, research shows that more content promotes trust (Gefen, 2000; Lee & Turban, 2001; Luhmann, 2000) and contributes to persuasion and ultimately conversion (Cialdini, 2009; Fogg et al., 2002; Li & Chatterjee, 2010). On the other hand, content was suggested to alienate users by increasing effort and friction (Geissler, Zinkhan, & Watson, 2006; Kahneman, 1973; Norton, Frost, & Ariely, 2007; Song & Schwarz, 2010; Vishwanath, 2004). As a result, the impact of the volume of content provided is still unsolved.

There is a growing interest in Information Systems (IS) research on the determinants of consumer behavior, from the way users evaluate products and services to the way they engage with online advertising (Smith & Anderson, 2016). This interest is motivated by the proliferation of new information and communications technologies (ICT), making commercial transactions over computer-mediated networks (e-commerce) to increasingly become the main form of conducting business. According to a recent survey, 79% of U.S. adults reported ever making an online purchase, spending nearly $350 billion annually (Smith, 2017). There is also an exponential growth in online marketing ("Carat Ad Spend Report," 2016).

To answer the research question, a pilot study (n= 535) and a series of large-scale online experiments (n= 27,083) were conducted in real-world commercial settings. Following an extensive literature review, two variations of landing pages, long and short, were created based on contemporary behavioral theories. Both pages had an identical form to collect users' information (a "conversion"). A between-group design was utilized to observe reactions to the page variations.

The rest of the paper is structured as follows: First, an extensive literature review is presented to identify relevant research streams and formulate a research question. The next sections introduce the research design, analyses and the explanation of results.

# THEORETICAL BACKGROUND

The decision-making processes that make users comply online - disclosure of personal information or purchase - have been researched in various domains. With regards to content, there are theoretical and empirical evidences on the influence of content on consumers' perceptions, intentions and eventual behavioral actions (Constantinides, 2004; Huizingh, 2000; Koufaris, Kambil, & LaBarbera, 2001; Nielsen & Loranger, 2006; Richardson, Dominowska, & Ragno, 2007; Wells et al., 2011). Previous research suggests that actual observable behavior is a reliable measure, as other aspects, like perception and intention, are directly related to it (Figueiredo, Almeida, Benevenuto, & Gummadi, 2014).

In the current case, the observable behavioral outcome is disclosure of personal information, operationalized in the form of users providing their e-mail (conversion). This behavior is described as consumer's willingness to rely on the seller and take action in circumstances where such action makes the consumer vulnerable to the seller (Luhmann, 2017). It is often conceptualized as "compliance", defined as "behavior change devoid of pressure" (Cialdini & Goldstein, 2002).

There is a seemingly inexhaustible battery of techniques to evoke compliance and conversion in IS literature (Benbunan-Fich & Koufaris, 2008; Gefen, 2000; Koufaris, 2002). However, the question of content volume is still unanswered, with various competing models attempt to provide a theoretical underpinning.

Studies that support more content emphasize that greater volume is needed to mitigate risk and communicate value (C. W. Chen & Koufaris, 2015; Eisingerich & Kretschmer, 2008; Tversky & Kahneman, 1985); whilst other studies stress that less content is crucial to minimize processing effort, focus users' attention and even generate positive affect (Alter & Oppenheimer, 2008; Brynjolfsson & Smith, 2000; Norton et al., 2007; Song & Schwarz, 2008). The following sections review the two streams of research and identify corresponding content features.

## IN FAVOR OF LONG-FORM (GREATER VOLUME OF CONTENT)

Research literature that supports greater volume (or amount) of content emphasizes that online commercial interactions involve a certain amount of risk (Lee & Turban, 2001; Lim, Sia, Lee, & Ben-basat, 2006; Luhmann, 2017). To mitigate it, web pages should display more content to generate trust, communicate value and signal quality (Eisingerich & Bell, 2008; Gefen, 2000; Luhmann, 2017; Rainie & Anderson, 2017).

Trust in a website is based on its design and content (Mavlanova, Koufaris, Benbunan-Fich, & Lang, 2015). Trust in an entity affects people's willingness to take action (Gefen, 2000; Luhmann, 2000, 2017). Trusting beliefs occur when a potential online shopper believes that the online store is benevolent, competent, honest, or predictable. Trust-building strategies include customer endorsements by similar (local, nonforeign) peers (Lim et al., 2006) and privacy and security policy. Although consumers may not mind the collection and use of their personal data, they would like to know how the data would be used (Bélanger & Crossler, 2011; Hargittai, Fullerton, Menchen-Trevino, & Thomas, 2010; Jarvenpaa et al., 2006). Online reviews and ratings were also found to be important: Most of the consumers consult online ratings and reviews when buying something for the first time (Gafni & Golan, 2016; Smith & Anderson, 2016).

Perceived value was also recognized as a factor of content. Research suggests that content is highly valued by online users and should convey information about the company, its products, and the level of service the consumer can expect to receive (Huizingh, 2000). The more useful information the website has, the more valuable it is for (Braun, Lee, Urban, & Hauser, 2009; Coker, 2013; Jennings, 2000; Wells et al., 2011). Signals that form a perception of value are achieved by informative content which includes the availability of customer reviews, shopping advice, articles, product information, and website policies. Good landing pages should explain the product or service offer and emphasize its value (Brynjolfsson & Smith, 2000). The basic premise of virtual experience is that if an organization can provide a consumer with website characteristics that afford a sense of telepresence (i.e., being there), consumers will be better able to evaluate the product, resulting in increased intentions to cooperate (Li & Chatterjee, 2010).

Content is also highlighted as a method for persuasion and influence. An established traditional theory is the Means-End Theory that suggests that to influence and persuade, the message should aim to lead the consumer to a desired end-state (Gutman, 1982). The six universal principles of social influence support to supplying information on authority, reciprocity, scarcity, social validation, likability and commitment and consistency. These principles serve as heuristic cues for decision making. When processing heuristically, individuals can use certain cues, rules of thumb, or surface features to determine whether to comply with a request (Cialdini, 2009). Research on credibility has also highlighted the need for elaboration, for example using online content to highpoint expertise or specify services (Fogg, 1998; Fogg et al., 2001; Fogg et al., 2002; Laja, 2014; Tseng & Fogg, 1999).

Signaling theory has been applied to digital content as a potential signal of product quality (Wells et al., 2011). Providing rich information represents an important signal, which in turn has a direct relationship with customers' purchase intention. Content signals may include product information, expert product reviews, press releases, frequently asked questions (FAQ), and news. Rich content on a website is one of the most vital signals influencing perceptions of the overall quality of the website, at the same time, not providing an appropriate valuable content signal of the seller's reluctance to invest time, effort and resources in providing information. The lack of detailed content may prompt buyers to believe that a seller has something to hide (Mavlanova et al., 2015). Online information can signal consumers about product or company quality, and whether this signal influences their willingness to transact with the company, and ultimately the prices they are willing to pay for the company's goods and services (Gregg & Walczak, 2008).

"Perceived value", "Trust", "Persuasion and influence", and "Quality" are often connected with long form, elaboration or greater content volume. For example, Gefen (2000) and Eisingerich and Bell (2008) provided empirical evidence that greater volume of information elicit familiarity and trust; Fogg et al. (2002) empirically demonstrated that elaboration and content volume are important fac-tors when aiming to persuade and influence consumers online; and Mavlanova et al. (2015) provided evidence that greater content volume signal quality and impact perceived value.

Supporting arguments for more content are summarized in Table 1.

### Table 1. Supporting arguments for more content

| Determinants | Content features | Supporting research |
|---|---|---|
| **Perceived value** | Presence of relevant information provided on the website. Convey intrinsic product attributes Written product features, pictures, and virtual product experiences. | (Gefen, 2000; Luhmann, 2017; Rainie & Anderson., 2017; Wells et al., 2011). |
| **Trust** | Trust in an entity affect people's willingness to take action. Content should include customer reviews, shopping advice, product information and the availability of website policies (e.g., privacy policy). | (Bélanger & Crossler, 2011; Hargittai et al., 2010; Jarvenpaa et al., 2006, 2006; Lee & Turban, 2001; Lim et al., 2006; Mavlanova et al., 2015) |
| **Persuasion and influence** | The message should aim to lead the consumer to the desired end-state. Content should supply information on authority, reciprocity, scarcity, social validation (also called social proof), likability, and commitment and consistency. | (Cialdini, 2009; Cialdini & Goldstein, 2002, 2004; Fogg, 2009) |
| **Quality** | Rich information signal quality, credibility, reputation and size. | (Gregg & Walczak, 2008; Mavlanova et al., 2015) |

## IN FAVOR OF SHORT-FORM (LESS CONTENT)

In contrary, some researchers suggest that effort is the real cost for users and advocate minimalist approach (Molich & Nielsen, 1990; Nielsen, 2005; Nielsen & Molich, 1990). High content volume signals effort while less content is easier to process (Norman, 2013). Perceived effort was suggested to be a barrier to conversion: 76% of consumers stated that the most important factor in a website's design is that it "makes it easy to find what I want" (Gofman, 2007).

Prior research suggests that when faced with complex and uncertain situations, individuals tend to use simple heuristics and cues in a bounded rational decision-making process and make relatively 'uninformed' judgments on the basis of a minimum of information (Brynjolfsson & Smith, 2000). Research shows that elaboration may result in "Friction," defined as a psychological resistance to a given element on the page (Lindgaard et al., 2006; Vishwanath, 2004). The volume of content and the control of users' attention are inseparable. It takes users less than a second on average to evaluate a website's appeal after viewing it for the first time (Kolko, 2015). The Hick–Hyman Law (Hick's law) describes the time it takes users to decide in light of the possible choices they have. When the user's attention is diverted to access necessary information, there is an associated cost in time or effort, which is called the "Information access cost." It can come at a price: if processing the information is too demanding, the working memory disengages and moves on (Miller, 1956).

Simple content is scientifically shown to be easier to process, but also leads to greater appreciation (Tuch, Presslaber, StöCklin, Opwis, & Bargas-Avila, 2012). Research showed that more information leads, on average, to less liking and dissimilarity (Norton et al., 2007).

Information overload is associated with a host of undesirable outcomes including diminished productivity, poor decision making, and negative perception (Y. C. Chen, Shang, & Kao, 2009; Lucian, 2014; Soto-Acosta, Jose Molina-Castillo, Lopez-Nicolas, & Colomo-Palacios, 2014). Information overload influences not only which information users choose to consume, but also how they consume it. Since much of the information available to users is increasingly abundant and immediately available, attention has become a limiting factor in the consumption of information (Davenport & Beck, 2001). As a result, information interactions are becoming highly asymmetrical, thus, only a small portion of information receives significant attention, while the remaining majority is barely noticed (Szabo & Huberman, 2008). In addition, less than a third of the information that is noticed is read by users (Nielsen, 2015). Research suggested that problems that can arise from information overload when individuals with limited cognitive abilities encounter massive amounts of potentially relevant information (Y. C. Chen et al., 2009; Eppler & Mengis, 2004).

Elaboration is also suggested to alienate users by closing their "Knowledge Gap" (also called "Information Gap"), defined as the difference between what users know and what they would like to know (Losee, 2012; Menon & Soman, 2002) . Curiosity is generated when a person becomes aware that a knowledge gap exists – they would be motivated to search for more information to close the gap (Menon & Soman, 2002). Supporting arguments for less content are summarized in Table 2.

**Table 2. Supporting arguments for less content**

| Determinant | Information factors | Supporting research |
| --- | --- | --- |
| Perceived effort/ease of use | Simple content is easier to process. If processing the information is too demanding, the ing memory disengages and moves on. | (Gofman, 2007; Nielsen, 2005; Nielsen & Molich, 1990; Norman, 2013; Norton et al., 2007; Tuch et al., 2012) |
| Attention | When the user's attention is diverted to access. Necessary information, there is an associated cost in time or effort. | (Lindgaard et al., 2006; Vishwanath, 2004) |
| Curiosity | Curiosity is generated when a person becomes aware that a knowledge gap exists (less information). | (Losee, 2012; Menon & Soman, 2002) |
| Information overload | Limited attention, poor decision making, and negative perception | (Y. C. Chen et al., 2009; Davenport & Beck, 2001; Eppler & Mengis, 2004; Lucian, 2014; Nielsen, 2015; Soto-Acosta et al., 2014; Szabo & Huberman, 2008) |

## THE RESEARCH QUESTION

The literature review has underlined a few gaps in current research:

(1) No clear inference on how the volume of content influences consumer behavior.
(2) Lack of agreement and opposing theories on the advantages or disadvantages of elaboration on commercial web pages.
(3) Most studies explore content features, yet not enough emphasis is on the volume of content.

(4) While there are many professional articles on the topic, there is a dearth of academic research that illuminates the role of content in landing pages, its influence on users' compliance or how it can be used for conversion optimization.

To begin addressing these gaps, this research explores whether the volume of content provided on a landing page (independent variable) impacts users' behavior (dependent variable). The variables were operationalized as the volume of content displayed to the users and users' conversion rate (their willingness to disclose personal information in the form of providing their e-mail). The hypothesis is that volume of content impacts behavior. According to the literature, there is no clear inference whether long-form or short-form performs better.

# METHODOLOGY

## DESIGN

The hypothesis was tested in a series of large-scale online experiments. All the experiments took place in commercial settings and focused exclusively on landing pages that appeared in response to clicking on an online advertisement. The purpose of the landing pages is to persuade a visitor to take action by completing a transaction, in this case, submission of an e-mail address (an action called "conversion").

A simple experimental design consisting of one control group and a treatment group was employed. All experiments were based on single-factorial A/B tests, also called "split testing." In an A/B test, two variations of the same page ("A" and "B") are created and differ only in the element that is being tested (for this study purpose, volume of content). Visitors were randomly assigned to the pages. External loading times, web addresses and other external factors were equalized. Doing so enabled content to be isolated as an independent variable and thereby to observe its direct influence on a behavioral action as the outcome of interest. This method allowed establishing a causal relationship between changes and their influence on user-observable behavior with high probability.

### Page variants

Two versions of landing pages were created: one with detailed information (control) and another version that seemed to be the same, only most of the information was removed (treatment), as presented in Figure 1. Both pages promoted a digital service - ClaimFame (www.claimfame.com), a marketplace that connected content creators with talent on a broad range of media projects. Page variants included an identical form at the top of the page prompting the users to sign-up for a newsletter by providing their e-mail address (observed outcome, "conversion"). The page versions differed only in volume of content provided: Variant A was based on theories of persuasion, trust, and signaling and included information on the service (perceived value), customer reviews (quality) and privacy policy (information on how the personal data provided will be used). Version B, on the contrary, was based on simplicity and processing fluency with no additional content besides the form (include no information on the service that the users are signing for or what will be done with their e-mail address). Version "B" was minimalistic and short, containing only 42 words, 211 characters, and 12 lines, with a simple call-to-action form; version "A" was identical to version "B" but added information about the service, making it longer - 278 words, 1,501 characters and 49 lines.

The page variants were sent to an external consulting company and a panel of marketing experts in order to verify their internal and external validity. The experts corroborated the validity of the pages' design and confirmed that they adequately represent variations in content volume as accepted in applied industry settings.

| Variant A: More content | Variant B: Less content |
|---|---|
| 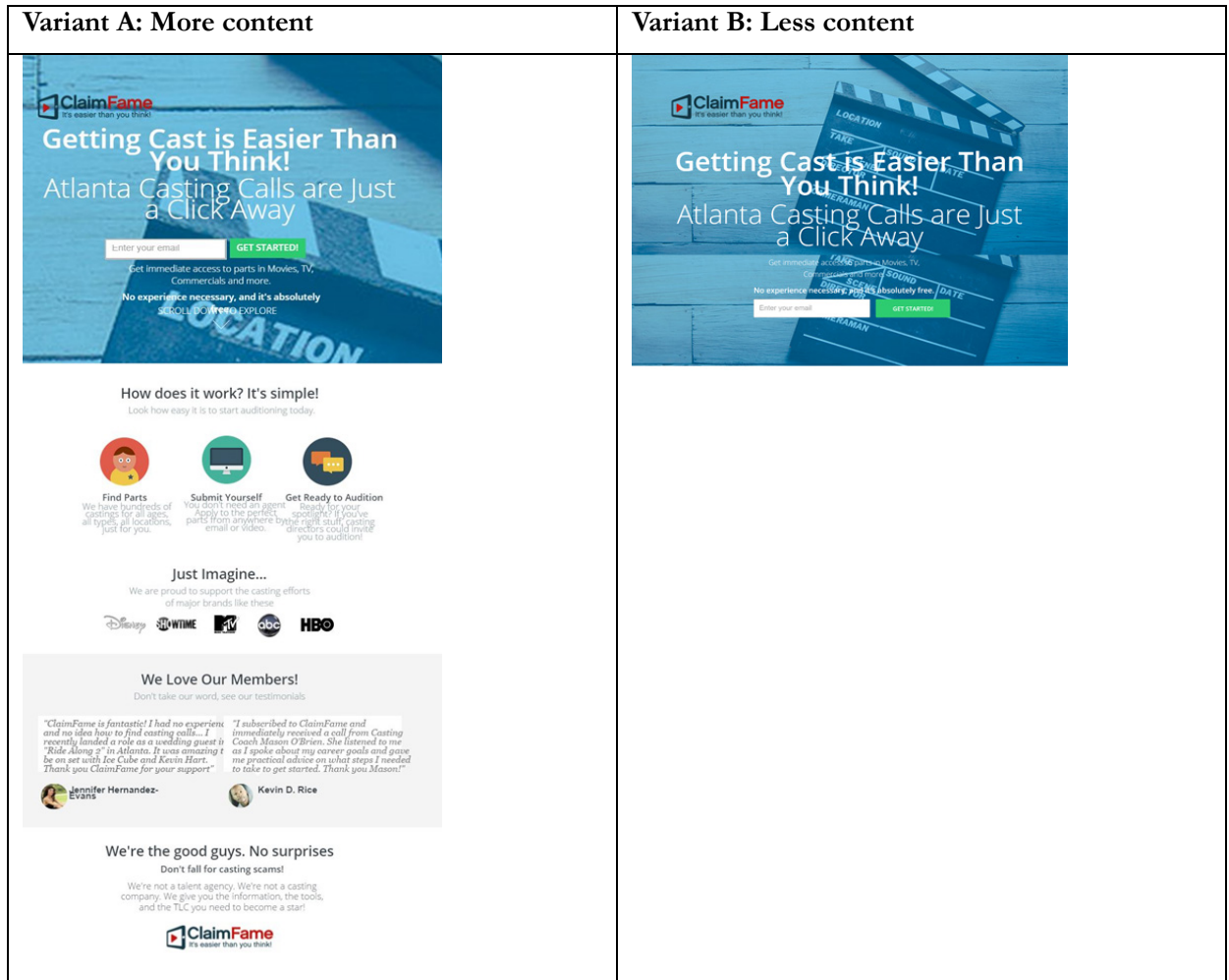 | |

**Figure 1: Page variations used for the experiments**

## Participants

Users' traffic to the page was generated using AdWords, a web-based advertising tool by Google, offering paid display advertisements in Google's search results based on relevant keywords (as shown in Figure 2). Google AdWords are increasingly used to recruit people into research studies. The service was proven to be a reliable method to recruit real-web users and to accurately control for location by targeting control areas (Jones, Goldsmith, Williams, & Kamel Boulos, 2012).

## Procedure and data collection

The pages were published online using Unbounce.com, a self-serve hosted service providing a suite of tools to create, publish and test landing pages. Unbounce.com was chosen because it was regarded as a market trailblazer, widely used, and conveniently provided relatively simple means to build landing pages and monitor their performance by built-in web analytics. Traffic is evenly split between them: 50% of the users are shown page "A", while the other 50% are taken to page "B". Unbounce.com used cookies to make sure users were always directed to the same page and that the split would be equal. If users landed on page "A", a cookie was placed on their computer so that even if they came back later, they would always see version "A". This was important to ensure that users would not notice the testing.

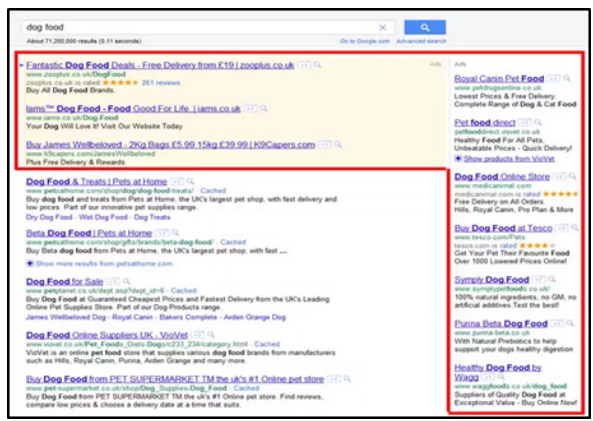

**Figure 2. Example of Google Adwords used to recruit real web users.**

Data was collected on all visitors from the site and included click-through rate (number of visitors), their IP address, location, date and time of the visit and of course number of users who provided their e-mail addresses (conversion tracking method).

## THE RESULTS WERE ANALYZED USING IBM SPSS VERSION 22. EXPLORATORY EXPERIMENT

An initial experiment was first set up to examine the viability of the design. Only a specific geographical location of users within the United States was targeted (Atlanta, Georgia), and set to stop the experiment when more than 200 user conversions were reached (Google's guidelines recommend at least 100 conversions per page before deciding which version is best). Overall, the number of unique visitors directed to the page variations was 535 (n=535).

### EXTENDED EXPERIMENTS

To better test the impact of content volume on consumer behavior, the exploratory experimentation was extended and a large nationwide, multi-market, split testing was conducted. Five new experiments were performed in the same manner with different populations. The experiment used the same landing page variations from the exploratory experiment: Version "A", with more content (long-form), and version "B", with minimal content (short-form).

The sample size was expanded by directing a greater volume of user traffic using Google AdWords. This time geo-targeting was utilized by targeting users in four specific regional US markets: Atlanta, Miami, Los Angeles, and New York. The experiment also maintained a fifth, "national" group that consisted of users directed from all over the United States. It is a well-known research method called "Blocking," which is based on the arrangement of experimental units into groups (blocks/lots) consisting of units that are similar to one another (Addelman, 1969). Blocking reduces sources of variation between units and thus allows greater precision in the estimation of the source of variation under study. In randomized block designs, there is one factor or variable that is of primary interest. However, there are also several other nuisance factors that may affect the measured result but are not of primary interest. In this case, hour and day (time) is an example of such a factor. Within blocks, it is possible to assess the effect of different levels of the factor of interest without having to worry about variations due to changes of the block factors, which are accounted for in the analysis.

The expanded experiments were conducted over a two month period. Overall, data was collected from 27,083 unique visits to the page variants.

# RESULTS

The results of the exploratory experiment, as shown in Table 3, demonstrate a clear advantage in conversion on the short form pages. While the results are not sufficient, they provide preliminary information for more definite investigation.

**Table 3. Summary of conversion results for exploratory experiment**

| Page Variant | Total unique visitors | Conversions | Users who didn't convert | Conversion rate |
|---|---|---|---|---|
| **"A" (Long)** | 273 | 106 | 167 | 38.83% |
| **"B" (Short)** | 262 | 140 | 122 | 53.44% |
| **Total** | 535 | | | |

As can be seen, the percentage of users who provided their e-mail was calculated as a percentage of the total amount of visitors to the page (conversion rate). Out of 535 visitors to both pages, on the long variant 38.83% of the users converted (provided their e-mail address). In comparison, in the short variant, the conversion was by 53.44% of the users. This is a significant 37.62% increase in conversion rate, rendering the short variant a clear winner. The results were statistically analyzed founding significant difference between the groups (t-value -3.6569, sig 0.00643, $p < 0.05$), and chi-square (11.4848, p-value is 0.000702, $p < 0.05$).

In the extended experiments, the overall data was collected from 27,083 unique visits to the page variants that resulted in 9,593 conversions.

**Table 4- Summary of conversion results for expanded experiments**

| Experiment / Campaign | Page variant | Total unique visitors | Conver-sions | Conversion Rate | t |
|---|---|---|---|---|---|
| **National** | "A" (Long) | 5497 | 1405 | 26% | -6.651** |
| | "B" (Short) | 5494 | 1718 | 31% | |
| **Atlanta** | "A" (Long) | 3690 | 1231 | 33% | -14.560** |
| | "B" (Short) | 3688 | 1838 | 50% | |
| **Miami** | "A" (Long) | 1235 | 444 | 36% | 2.470** |
| | "B" (Short) | 1237 | 639 | 52% | |
| **Los Angeles** | "A" (Long) | 1945 | 529 | 27% | -13.817** |
| | "B" (Short) | 1948 | 937 | 48% | |
| **New York** | "A" (Long) | 1176 | 324 | 28% | -8.946** |
| | "B" (Short) | 1173 | 528 | 45% | |
| **Total** | "A" (Long) | 13543 | 3932 | 29% | -22.166** |
| | "B" (Short) | 13540 | 5660 | 42% | |
| | Both | 27,083 | 9,593 | | |

** $p < 0.01$

As seen from Table 4 and in Figure 3, while a similar amount of users received the "A" or "B" variant in each market, the short variants significantly outperformed the long one across all markets. T-tests were performed also for the percent of conversions in "A" version groups for all the campaigns (t= -20.759, df= 119, sig=.000) and for the percent of conversions in "B" version groups for all campaigns (t= -20.131, df= 119, sig=.000).

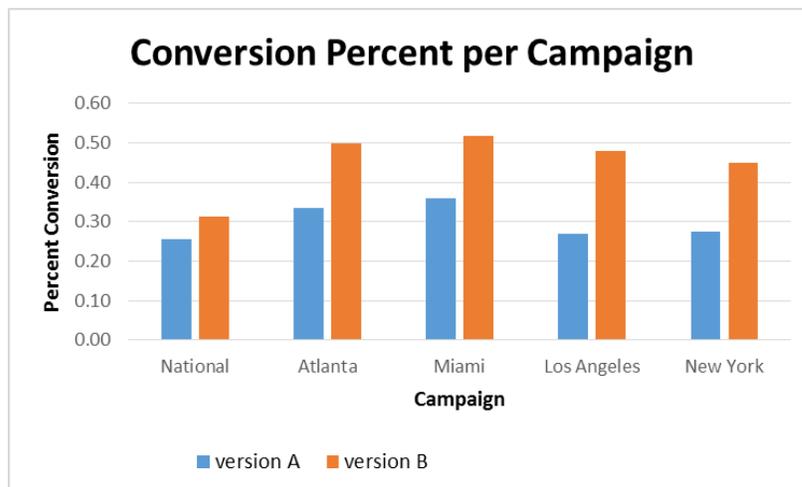

**Figure 3. Comparison of "A" and "B" conversion percent for each campaign**

## Location and time

The results also show some influence of the location on users' conversion rate. As can be seen from Figure 3, there are differences between the various campaigns. In Miami, for example, people tend to convert more than other places, for both versions. In the National campaign, both conversion rates are minimal. In Los Angeles, the disparity between the conversion of the "A" and "B" versions is the greater while in the National campaign these are very similar.

While not significant, the fact that the percentage of conversions differed between locations must be further examined.

The influence of the time users visited the page was also checked. According to the time-stamp of each visit, the day of the week of each visit was computed (Sunday-Saturday). The day of the week may be an interesting parameter, to understand if leisure time affects the user decision of conversion.

No differences were found between the day of the week and the average conversion of each page, for the sum of all experiments, as can be seen in Figure 4. T-test was performed between the average conversions for the "A" version for the different days in the week (t=10.805, df=34, sig=.000, p<0.005), and same for the "B" version (t=10.341, df=34, sig=.000, p<0.005). No statistically significant correlations were found. The average conversions for both pages are similar during all week, and the rate between the long page and the short one is similar in each day of the week. No statistical differences were found when checking each campaign.

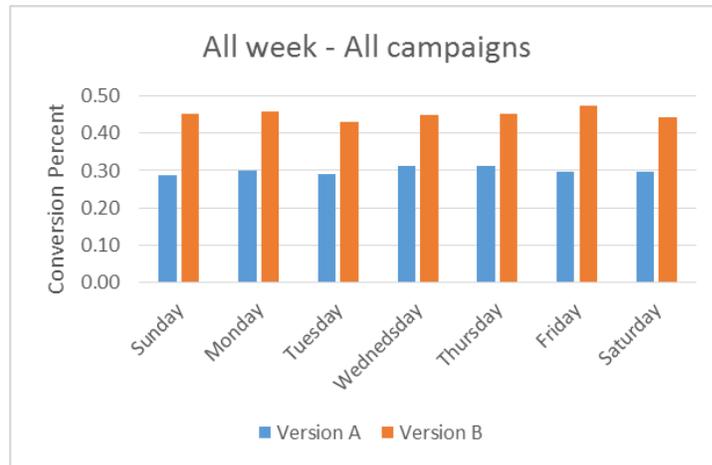

**Figure 4. Conversion Percent for all campaigns together, all week along**

Next, the hour of the visits was explored. Based the time-stamp the hour of the day was extracted. The time-stamp was originated in New-York time zone. New York, Atlanta, and Miami correspond to the same time zone. To arrange the hour for Los Angeles visits, the time was reduced to 3 hours, thus, defining the "real" hour. The National campaign was omitted from this examination because the data was collected from all around the USA, with different time zones.

T-tests were performed to compare the "A" and "B" versions conversions for each hour. For all four campaigns together, for most of the hours, statistical differences were found between the campaigns.

Comparing each hour for the "A" version, between all four campaigns, there were no statistical differences in population behavior. The same result was observed for the "B" version.

The conversion rates between A and B versions across all 24 hours was examined. As can be seen in Figure 5 there were differences at the 5:00 and 22:00 times. These times can be associated with periods around waking up and going to bed.

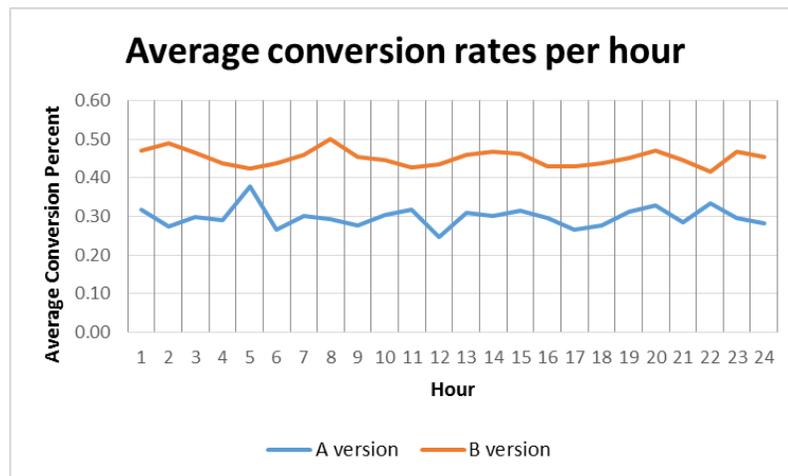

**Figure 5. Average conversion rates for all campaigns together, per hour**

## DISCUSSION

Results show a direct correlation between the volume of content on a web page and users' decision-making which lead to conversion. Surprisingly, the fewer information users had, the more they were inclined to provide their personal data. The results show a significant advantage to the shorter pages.

The short landing pages had higher conversion rates across locations, days and hours, and consistently outperformed their longer equivalents.

These results have a few interesting implications:

**Content impact behavior.** First, an intuitive conclusion is that content is a crucial determinant of consumers' behavior. Since the research method allowed for the effects of intention and other variables to be controlled, volume of content was isolated as the only factor differentiating between the pages. The differences in the behavioral outcome show that consumers' behavior, especially conversions, hinges on the nature of the content being consumed.

**Less content elicits conversions (desired consumer behavior).** Second, the results show a clear advantage to using less content on landing pages. In all of the experiments, the pages with less content had significantly better percentages of the desired outcome. This finding supports the research stream arguing for less content as a determinant of conversion.

**Trust or ease-of-use?** Third, results suggest an interesting observation about the nature of online consumers' behavior. As seen, users that did not have any information on the service, its value or how their personal data will be used tended to convert more. By contrast, users who were exposed to this information were less inclined to convert. This is in contradiction to theories of trust, quality and perceived value discussed earlier. It shows that mitigating risks and conveying value are not always important factors in determining consumers' willingness to take action in circumstances where such action makes the consumer vulnerable to the seller.

This finding suggests that online decisions, such as providing personal information on commercial web pages, are not always pragmatic. In these experiments, the *less* users were informed the more they were inclined to cooperate. It is, therefore, possible that individuals may make relatively 'uninformed' judgments on the basis of a minimum of information, without engaging in any form of deep cognitive and conscious reflection.

**Related factors (time, location).** Fourth, results showed some variations relating to the location of the users and the time of the visit to the page. While not statistically significant, the fact that the percentage of conversions differed between locations and times indicate that content is not the only factor influencing behavior. A possible hypothesis is that users' behavior differed in locations because of demographics (age, gender), economic status or personal preferences. The time of day can also be a factor influencing the rate of conversion. For example, trying to find differentiation in behavior during the day was interesting. It was found that at 5:00, when the people wake up, and at 22:00 at the end of the day, people tend to convert in the long version ("A") more than in other time of the day. This may suggest that users have more attention span in the early morning and at night, while during the day they prefer less content.

These results provide empirical evidence on the extent to which *content by itself* determines consumers' behavior. This research also contributes to the ongoing discussion on the type of information necessary to encourage website users and potential customers to take action.

While previous studies show that more content promotes trust (Gefen, 2000; Lee & Turban, 2001; Luhmann, 2000) and contributes to conversion (Cialdini, 2009; Fogg et al., 2002), this research shows otherwise. It supports prior work that suggests that increasing volume of content alienates users (Kahneman, 1973; Norton et al., 2007; Vishwanath, 2004).

The findings add some controversy to the relationship between the volume of content and consumers' behavior as they partially counter other studies. They also contribute to the scholarly literature on landing pages and online behavior.

# CONCLUSION

The findings demonstrate the extent to which the volume of content by itself determines consumers' behavior. The conventional wisdom of the web suggests that more content elicit consumer knowledge and persuasion, which ultimately leads to more desirable behavioral goals. Yet, this study has found the opposite can be true. Less content was actually proved to be a positive determinant of consumer behavior, while not enough to establish a linear relationship, results suggest that "less is more" can go a long way in creating commercial value.

This research highlights the importance of content on commercial landing pages and its influence on users' decision and actions. The results of this study show that content has a significant role in forming buyer's intentions and eventually willingness to provide personal data.

To the best of our knowledge, content volume and behavior on landing pages are rarely studied in isolation. This work falls within the larger vision of determining a method for effective communication on the web, which is of paramount importance in many theoretical and applied domains. Therefore, this research is unique and significant by focusing on the volume of content and landing pages, as the interplay between these factors is still highly debatable and under-explored.

# THEORETICAL AND PRACTICAL CONTRIBUTIONS

This study makes theoretical, methodological and practical contributions.

At a practical level, the results provide an amprical support in favor of less content as a determinant to desired consumer behavior. The results are informative for website designers, content strategists and marketers regarding the importance of content volume in commercial websites and production of more favorable purchase behavior intentions. The results from these studies have implications for most businesses using the e-commerce marketing channels and may contribute to conversion rate optimization. In addition, they can be used to as groundings for methodological content strategy.

At the theoretical level, the findings provide evidence on the importance of digital content in consumer behavior research. They contribute to IS research as well as other disciplines with regards to the determinants of users online and advances current knowledge of how content and consumers interact in online commerce. Specifically, results suggest that less content is sometimes better. This study supports previous research on the paradoxical relationship between the increased level of information and online decision making emphasizing the value of minimizing content. The empirical evidence suggests that contrary to previous work, not all behavioral and psychological theories are effective online, e.g., the well-established frameworks of persuasion (Cialdini & Goldstein, 2002; Fogg, 2009) and trust (Gefen, 2000; Luhmann, 2017).

Using this study as a starting point, it set the foundation for more exploration on how lack of knowledge paradoxically leads to higher conversions and overall user compliance. This research can argue in favor of the persuasive power of ambiguity and simplicity. The framework developed in this paper can serve as the basis for further studies on conversion and persuasion in online settings and other domains.

# LIMITATIONS AND FURTHER RESEARCH

As in every behavioral study, a major limitation of this study is a difficulty to identify clear causation in human behavior. There are various factors that illicit cognitive and emotional responses of end-users; therefore, it is challenging to assess clear relationship due to the number of factors involved and the study's interdisciplinary nature. The claim that "less is more" is grounded on the assumption of a linear relationship. However, one might suggest that this assumption might not be true if the form is too short. Another limitation is that the results may not be generalized/applicable to other domains. The contested issue with regards to field experiments is their external validity. Given that field experiments necessarily take place in a specific geographic and political setting, the extent to which findings can be extrapolated to formulate a general theory regarding economic behavior is a concern.

It is important to remember that the findings presented here are limited and preliminary. There is a need to analyze additional data sets further, as well as examine the influence of other variations in design elements and user demographics.

Using this research as a starting point, future research can examine the impact of content in other contexts. It is also interesting to explore the reasons for consumer behavior beyond content, by focusing on demographic data and comparing it to observed behavior. Future work may include conducting a study that shows a connection between the two designs and persuasion as well as perceived value, trust, quality, perceived effort, ease of use, liking, attention, and curiosity. It is also possible to do so in a two-stage design (a quantitative or qualitative study and then an experiment). Finally, there is still no clear answer regarding which content features have the most influence on users, or what is the ideal volume of content. It will be valuable to explore other predictors, or relevant if it is possible to identify specific content features that are more influential than others. Future work should explore how to derive reliable and reusable metrics and methodologies that will inform the larger task of evaluating, developing and creating engaging content.

# BIOGRAPHIES

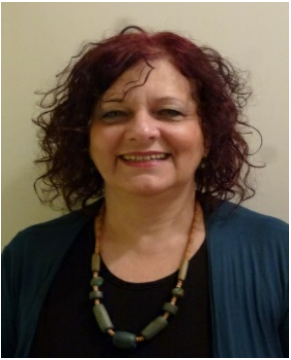

**Ruti Gafni** is the Head of the Information Systems B.Sc. program at The Academic College of Tel Aviv Yaffo. She holds a PhD from Bar-Ilan University, Israel (in the Business Administration School), focusing on Information Systems, an M.Sc. from Tel Aviv University and a BA (Cum Laude) in Economics and Computer Science from Bar-Ilan University. She has more than 40 years of practical experience as Project Manager and Analyst of information systems.

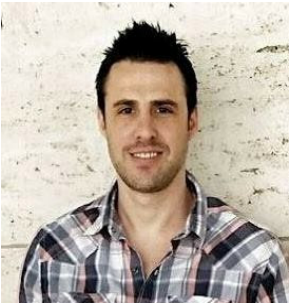

**Nim Dvir** is a Ph.D. student in the Information Studies department at the University at Albany. His main research interests are human-computer interaction (HCI), user experience (UX), E-Commerce, Content Strategy, and online user behavior. His work explores the determinants of perception, motivation, engagement, and decision-making in digital environments. He holds a B.A in International Relations and Media Studies from New York University (NYU) and an M.B.A in Marketing and Information Systems from City University of New York (CUNY) Baruch College's Zicklin School of Business, both Cum Laude. For more information: albany.edu/~nd115232/